\documentclass[10pt,twocolumn]{article}

\usepackage[utf8]{inputenc}
\usepackage[T1]{fontenc}
\usepackage[english]{babel}
\usepackage{booktabs} 
\usepackage{makecell}
\usepackage{multirow} 
\usepackage{xcolor}
\usepackage{colortbl}
\usepackage{comment}
\usepackage{titlesec}
\usepackage{array, booktabs}
\titlelabel{\thetitle.\quad}
\usepackage{algorithm}
\definecolor{blue}{HTML}{2197ff}
\colorlet{rowblue}{blue!8!white}  

\definecolor{rowgray}{gray}{0.95}
\usepackage{geometry}
\usepackage{authblk}
\usepackage{amsmath, amssymb}

\usepackage{graphicx}
\usepackage{caption}
\usepackage{subcaption}
\usepackage{algpseudocode}
\usepackage{adjustbox}
\usepackage[most]{tcolorbox}
\usepackage[square,numbers]{natbib}
\usepackage{hyperref}
\hypersetup{
    colorlinks=true,
    linkcolor=blue,
    citecolor=blue,
    urlcolor=blue,
}

\tcbset{
  myquote/.style={
    enhanced,
    colback=white,
    colframe=white,
    borderline west={2pt}{0pt}{gray},
    boxrule=0pt,
    sharp corners,
    left=6pt,
    right=0pt,
    top=0pt,
    bottom=0pt,
  }
}

\usepackage{xcolor}
\usepackage{caption}

\definecolor{green}{cmyk}{0.24, 1.0, 0.78, 0.07}

\DeclareCaptionLabelFormat{figlabel}{\color{green}$\blacktriangledown$
\textbf{#1~#2}}
\DeclareCaptionLabelFormat{tablabel}{\color{blue}$\blacktriangle$ \textbf{#1~#2}}

\title{
  \rule{\linewidth}{1pt} \\[0.5ex]
  \textbf{\bfseries Interlaced dynamic XCT reconstruction with spatio-temporal implicit neural representations}
  \rule{\linewidth}{1pt}
}
\author[1, 2]{Mathias Boulanger}
\author[1, 3]{Ericmoore Jossou}
\affil[1]{Nuclear Science and Engineering, Massachusetts Institute of Technology, 77 Massachusetts Ave, Cambridge, 02139, Massachusetts, United States}
\affil[2]{École Polytechnique de Bruxelles, Université Libre de Bruxelles, 50 Franklin Roosevelt Ave, 1050, Bruxelles, Belgium}
\affil[3]{Electrical Engineering and Computer Science, Massachusetts Institute of Technology, 77 Massachusetts Ave, Cambridge, 02139, Massachusetts, United States}
\date{\today}

\begin{document}

\maketitle

\begin{abstract}
In this work, we investigate the use of spatio-temporal Implicit Neural Representations (INRs) for dynamic X-ray computed tomography (XCT) reconstruction under interlaced acquisition schemes. The proposed approach combines ADMM-based optimization with INCODE, a conditioning framework incorporating prior knowledge, to enable efficient convergence. We evaluate our method under diverse acquisition scenarios, varying the severity of global undersampling, spatial complexity (quantified via spatial information), and noise levels. Across all settings, our model achieves strong performance and outperforms \textit{Time-Interlaced Model-Based Iterative Reconstruction} (TIMBIR), a state-of-the-art model-based iterative method. In particular, we show that the inductive bias of the INR provides good robustness to moderate noise levels, and that introducing explicit noise modeling through a weighted least squares data fidelity term significantly improves performance in more challenging regimes. The final part of this work explores extensions toward a practical reconstruction framework. We demonstrate the modularity of our approach by explicitly modeling detector non-idealities, incorporating ring artifact correction directly within the reconstruction process. Additionally, we present a proof-of-concept 4D volumetric reconstruction by jointly optimizing over batched axial slices, an approach which opens up the possibilities for massive parallelization, a critical feature for processing large-scale datasets.
\end{abstract}

\section{Introduction}
The tomographic observation of in situ dynamic phenomena such as melting and solidification of alloys is inherently ill-posed, as each sinogram encodes information from different temporal states of the evolving object. Traditional reconstruction techniques typically fail to recover high-quality time-resolved volumes under such conditions. At best, they yield temporally blurred reconstructions; at worst—when the object changes too rapidly—reconstruction becomes entirely unreliable, making it difficult to observe the microstructural changes. In this work, we explore the use of \textit{implicit neural representations} (INRs) as a flexible framework to address this challenge. INRs represent continuous signals via neural networks, enabling images to be reconstructed directly from spatio-temporal coordinates \cite{Where}. This formulation offers several key advantages: resolution independence, low data storage memory footprint, inherent smoothness, and the ability to generalize beyond discrete grid structures~\cite{Where}. These properties make INRs particularly attractive for tackling challenging inverse problems such as dynamic X-ray computed tomography (XCT) reconstruction from interlaced measurements.

Early work addressing dynamic tomography under severe sampling constraints proposed interlaced acquisition schemes that distribute projection angles across time, enabling the reconstruction of time-resolved volumes from a limited number of views per frame. For instance, Mohan et al.~\cite{TIMBR} introduced \textit{Time Interlaced Model-Based Iterative Reconstruction} (TIMBIR), a framework that combines interlaced angular sampling with 4D \textit{Model-Based Iterative Reconstruction} (MBIR), explicitly modeling the sensor noise statistics and detector non-idealities such as zingers and ring artifacts. By redistributing projection angles more uniformly in time, TIMBIR significantly increases temporal resolution without compromising spatial quality, outperforming traditional analytic reconstructions~\cite{TIMBR}.

A complementary strategy was introduced by Zang et al.~\cite{ST}, who proposed a space-time tomographic framework capable of jointly reconstructing volumetric sequences and estimating dense deformation fields from X-ray projections acquired under continuous object motion. Using an interlaced low-discrepancy angular sampling scheme together with a multi-scale alternating optimization, their method aggregates information across time while compensating for inter-frame deformation, enabling high-quality reconstructions for slowly and smoothly varying shapes.
Building on this work, Zang et al.~\cite{WARP} extended the formulation to handle high-speed, non-periodic deformations through a warp-and-project reconstruction model. This approach replaces the discretized time axis with a continuous one, assigns each projection a precise capture time, and enforces projection–volume consistency via forward and backward projection of the keyframes. The method, which also employs interlaced acquisition, adaptively inserts additional keyframes in periods of rapid motion, decouples acquisition and reconstruction frame rates, and significantly improves temporal fidelity and spatial accuracy compared to the original space-time tomography.

\subsection{Implicit Neural Representations for XCT Reconstruction}

Early neural implicit approaches such as CoIL~\cite{COIL} learn a continuous mapping from measurement coordinates to sinogram values, enabling dense and high-fidelity projection fields to be synthesized from sparse or irregular acquisitions, which can then be processed with standard reconstruction algorithms. IntraTomo~\cite{intratomo} later pioneered the direct use of INRs for tomography by parameterizing the attenuation field as a continuous coordinate-based multilayer perceptron trained directly from projection data via differentiable ray sampling. This formulation enables high-resolution reconstructions without voxel discretization, offering memory efficiency and inherent support for arbitrary sampling patterns. NeuralCT~\cite{neural} extended the direct INR approach by representing the volume as a continuous signed distance function and jointly reconstructing geometry and motion through differentiable projection operators combined with spatiotemporal regularization, enabling high-resolution, motion-compensated reconstructions from standard CT sinograms.

More recently, Neural Attenuation Fields~\cite{NAF} adapted INR-based tomography to sparse-view CBCT via self-supervised optimization of a continuous attenuation field using a multi-resolution hash encoding~\cite{mueller2022instant}, achieving high-fidelity reconstructions with reduced training times compared to earlier INR methods. A similar approach has been extended to dynamic or 4D CT settings. Notably, the work of Reed et al.~\cite{4D} introduces a 4D reconstruction pipeline that couples INRs with parametric motion fields to enable spatiotemporally continuous reconstructions from extremely limited angular coverage, using a differentiable Radon transform in a training-free, self-supervised framework. This method illustrates the potential of INRs not only for high-quality static reconstruction but also for handling complex temporal dynamics and motion. In a different line of work, non-periodic dynamic CT, especially for correcting non-periodic, rapid motion such as high-heart-rate cardiac imaging, has been addressed by BIRD~\cite{BIRD}, which employs a backward-warping implicit neural representation combined with diffeomorphic regularization to mitigate motion artifacts.

These methods collectively demonstrate the flexibility of INRs in XCT reconstruction tasks, ranging from sparse static recovery to dynamic and limited-angle scenarios. However, in dynamic settings, existing approaches typically assume that motion can be represented as a continuous, topology-preserving transformation, often parameterized via a diffeomorphic motion field or implemented through backward-warping. Such formulations inherently prevent modeling non-topological changes, such as the merging or splitting of structures, which traditional deformation models cannot capture. In contrast, our work aims to relax this assumption while leveraging the intrinsic strengths of interlaced acquisition scheme.

\section{Problem Formulation}
\label{sec:problem}

\subsection{Interlaced acquisition}
\noindent The conventional XCT acquisition strategy is progressive view sampling, in which projections are acquired continuously as the sample rotates at constant speed. A full set of angular views is then grouped and used to reconstruct a single static volume. In this configuration, the temporal resolution is dictated by the number of projections \( N_\theta \) required for accurate spatial reconstruction. If the system acquires projections at a frequency \( F_c \), the achievable volume frame rate becomes:
\[
F_s = \frac{F_c}{N_\theta}
\]

This linear relationship highlights a key bottleneck: increasing temporal resolution by reducing \( N_\theta \) reduces the spatial resolution, while increasing \( N_\theta \) reduces the frame rate. To mitigate this limitation, the interlaced acquisition strategy has been introduced~\cite{TIMBR}, enabling improved temporal coverage without increasing the total number of projections.

\begin{figure*}[!h]
    \centering
    \includegraphics[width=1\linewidth]{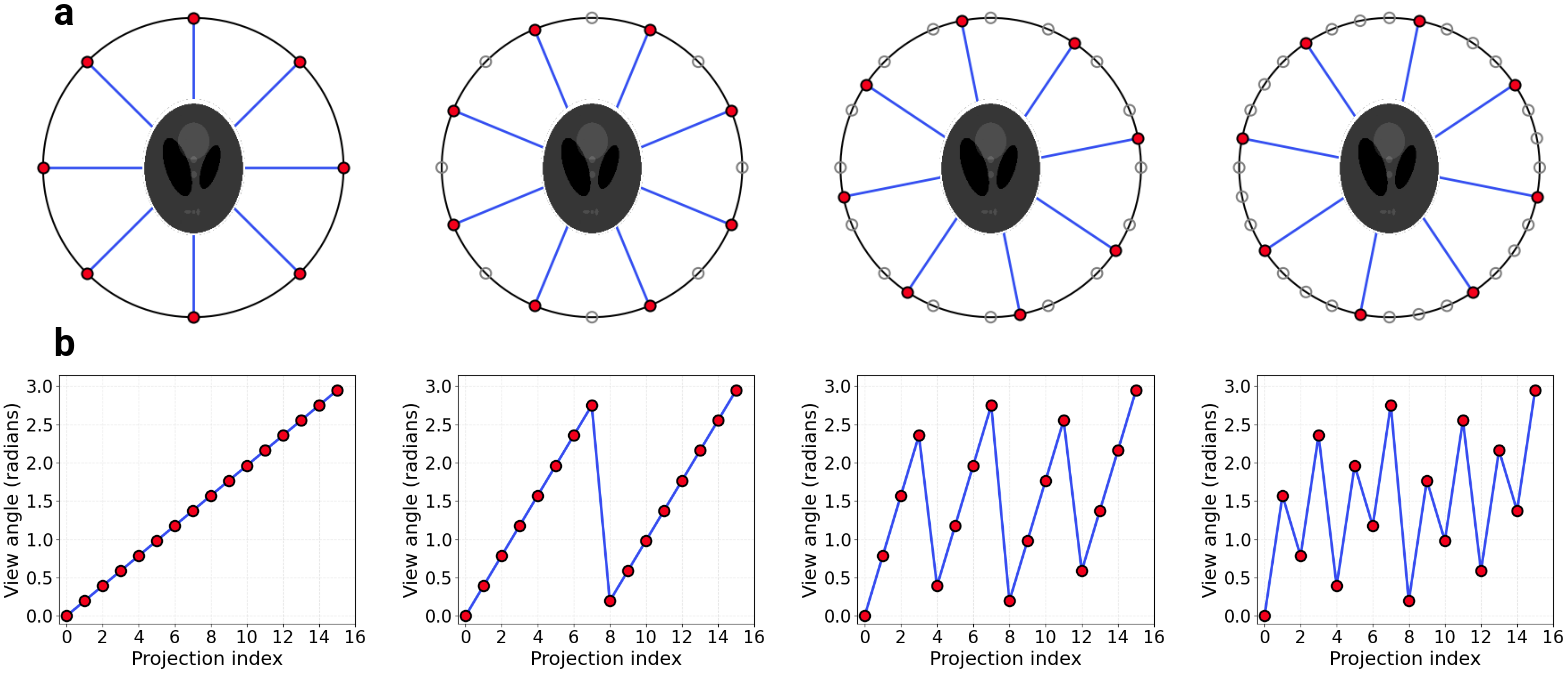}
     \caption[Interlaced acquisition]{(a) Visual representation of interlaced acquisition for K = 4 and $N_{\theta}=16$ . In practice, it is the sample that rotates. (b) Illustration of interlaced view sampling for different values of K and $N_{\theta}=16$. From left to right: K = 1, K = 2, K = 4 and K = 16.}
    \label{fig:interlaced_scheme}

\end{figure*}

In interlaced acquisition, the angular views are no longer grouped sequentially for each time frame. Instead, the full set of projection angles is distributed across multiple temporal frames in a staggered pattern. Each frame thus receives a sparse, non-contiguous subset of angular views, carefully selected to maximize angular diversity over time. This approach ensures that projections are spread as evenly as possible across both angle and time, which helps reduce aliasing and improve temporal sampling coverage. Over multiple frames, all angular views are still required to globally agree with the spatial Nyquist theorem, but each individual frame is undersampled from a spatial perspective. A typical interlacing pattern uses a bit-reversal sequence to determine the view assignment. For interlaced sampling over $\pi$ radians, the angle $\theta_n$ for the projection index $n$ is defined as \cite{TIMBR}:

\begin{equation}
\theta_n = \left[ \left( n \bmod \frac{N_\theta}{T} \right) T + \mathrm{Br}\left( \left\lfloor \frac{nT}{N_\theta} \right\rfloor \bmod T \right) \right] \cdot \frac{\pi}{N_\theta},
\label{eq:TIMBIR_eq}
\end{equation}
where $N_\theta$ is the total number of distinct projection angles, $T$ is the number of sub-frames, and $\mathrm{Br}(\cdot)$ is the bit-reversal over $\log_2 K$ bits.

The goal of this scheme is to increase the acquisition frequency to $F_s = T \frac{F_c}{N_{\theta}}$. Figure~\ref{fig:interlaced_scheme} illustrates the principle of interlaced acquisition. The benefit of interlacing lies in its ability to decouple temporal resolution from the total number of projections, without increasing acquisition speed. However, this gain comes at the cost of incomplete angular coverage per frame, which poses significant challenges for traditional analytic reconstruction methods.

\subsection{Static XCT scenario}
In a static XCT scenario, suppose we are given sinogram measurements $\mathbf{y} \in \mathbb{R}^{N_\theta \times N_d}$, where $N_\theta$ denotes the number of projection angles and $N_d$ the number of detector elements. Let $f_\theta : \mathbb{R}^n \rightarrow \mathbb{R}$ be an implicit neural representation (INR) parameterized by $\theta \in \mathbb{R}^p$, with $p$, the number of parameters. Let $\mathbf{x} = \mathcal{T}(f_\theta(\mathcal{G}))$ denote the image obtained by evaluating $f_\theta$ over a spatial grid $\mathcal{G} \subset \mathbb{R}^n$, discretized as a set of $H \times W$ coordinates. The goal is to learn the parameters $\theta$ such that the forward projection of $f_\theta$ matches the observed sinogram, i.e., $ P\{ \mathbf{x} \} \approx \mathbf{y}$, where $P$ denotes the forward Radon transform, which computes discrete approximations to the line integrals of $\mathbf{x}$ along the ray paths defined by the scanner’s acquisition geometry. This translates into the following optimization problem:
\begin{equation}
    \min_\theta \; \mathcal{L}(P\{\mathbf{x}\}, \mathbf{y}),
     \label{eq:loss1}
\end{equation}
where $\mathcal{L}$ is a differentiable loss function, typically a mean squared error.

\subsection{Dynamic XCT Scenario}

In the dynamic XCT scenario, we consider a sequence of sinogram measurements denoted by \( \mathbf{Y} = [\mathbf{y}_0, \dots, \mathbf{y}_{T-1}] \), where \( T \) is the number of temporal acquisitions. Each acquisition corresponds to a full gantry rotation, performed using an interlaced sparse sampling scheme.

We define the angular sampling strategy more precisely. Let \( \Theta = \{\theta_1, \dots, \theta_{N_\theta}\} \) denote the full set of projection angles. For each time step \( t \in \{0, \dots, T{-}1\} \), we acquire a subset \( \Theta_t \subset \Theta \), with \( |\Theta_t| \ll |\Theta| \), such that:
\begin{equation}
    \bigcup_{t=0}^{T-1} \Theta_t = \Theta, \quad \text{and} \quad \Theta_t \cap \Theta_{t'} \approx \emptyset \text{ for } t \ne t'.
\end{equation}

At each time step \( t \), we model the dynamic volume using a time-dependent implicit neural representation (INR), \( f_\theta(\cdot, t) : \mathbb{R}^n \rightarrow \mathbb{R} \). The reconstructed image is obtained by evaluating this representation over a spatial grid \( \mathcal{G} \subset \mathbb{R}^n \):
\begin{equation}
    \mathbf{x}_t = \mathcal{T}(f_\theta(\mathcal{G}, t)).
\end{equation}

Let \( \mathbf{X} = [\mathbf{x}_0, \dots, \mathbf{x}_{T-1}] \) be the reconstructed sequence, and \( \boldsymbol{\Theta} = \{\Theta_0, \dots, \Theta_{T-1}\} \) the corresponding angular subsets.

The reconstruction objective is formulated as the following optimization problem:
\begin{equation}
    \min_\theta \; \mathcal{L}\left(P_{\boldsymbol{\Theta}}\{\mathbf{X}\}, \mathbf{Y} \right),
    \label{eq:loss2}
\end{equation}
where \( P_{\boldsymbol{\Theta}} \) denotes the time-wise application of the Radon transform over the sequence of images, restricted to the respective angular subsets.

\subsection{Spatio-temporal regularization}

To promote both spatial and temporal smoothness in the reconstructed sequence, we incorporate total variation (TV) regularization terms in both space and time domains \cite{TV}.

The spatial TV encourages piecewise-smoothness within each frame, independently across time:
\begin{equation}
    \mathrm{TV}_\mathrm{space}(\mathbf{X}) = \frac{1}{T} \sum_{t=0}^{T-1} \left\| \nabla \mathbf{x}_t \right\|_1.
\end{equation}

The temporal TV penalizes abrupt variations over time by promoting smooth transitions between consecutive frames:
\begin{equation}
    \mathrm{TV}_\mathrm{time}(\mathbf{X}) = \frac{1}{T-1} \sum_{t=0}^{T-2} \left\| \mathbf{x}_{t+1} - \mathbf{x}_t \right\|_1.
\end{equation}

The final optimization problem is formulated as:
\begin{align}
 \min_\theta \; \mathcal{L}\left(P_{\boldsymbol{\Theta}}\{\mathbf{X}\}, \mathbf{Y} \right) 
 &+ \lambda_s \, \mathrm{TV}_\mathrm{space}(\mathbf{X}) \notag \\
 &+ \lambda_t \, \mathrm{TV}_\mathrm{time}(\mathbf{X}),
\end{align}
where \( \lambda_s \) and \( \lambda_t \) are regularization weights controlling the strength of spatial and temporal smoothing, respectively.

\section{Implementation}

\subsection{Encoding}
\noindent A central limitation of implicit neural representations (INRs) is their spectral bias, the tendency to favor low-frequency functions, which impairs their ability to capture high-frequency details~\cite{Where, hanin2019complexity}. A common remedy is positional encoding, which maps input coordinates to higher-dimensional spaces via sinusoidal or randomized projections to expand frequency capacity~\cite{mueller2022instant, tancik2020fourier}. For instance, a basic Fourier feature like $\gamma(x) = [\cos(2\pi x), \sin(2\pi x)]^T$ introduces periodicity. More advanced encodings of the form $\gamma(x) = [x, \dots, \cos(2\pi \omega^{j/m}x), \sin(2\pi \omega^{j/m}x)]^T$, with $\omega$ a frequency hyperparameter and $m$ the embedding dimension, further increase expressiveness~\cite{Where}.

While effective, these encodings are task-agnostic and fixed, lacking adaptation to signal structure. To address this, Implicit Neural Conditioning with Prior Knowledge Embeddings (INCODE)~\cite{kazerouni2023incode} introduces a conditioning mechanism that modulates the network's frequency response based on prior knowledge.

INCODE departs from traditional encodings by integrating a latent representation, extracted via pretrained models, that controls the network behavior. It employs a dual-network design: a \emph{composer} MLP with generalized sinusoidal activations, and a \emph{harmonizer} that predicts activation parameters from a latent code. The resulting activation function is:
\begin{equation}
    f(x) = \mathbf{a} \cdot \sin(\mathbf{b} \omega_0 x + \mathbf{c}) + \mathbf{d},
\end{equation}
where $\mathbf{a}, \mathbf{b}, \mathbf{c}, \mathbf{d}$ respectively control the amplitude, frequency, phase, and offset. These are dynamically predicted by the harmonizer from a pretrained encoder such as ResNet, enabling task-adaptive modulation of the frequency response.

In our experiments, we evaluated several encoding strategies to enhance the representative capacity of the INR, including Fourier Feature Mapping~\cite{tancik2020fourier}, Instant-NGP-style encodings~\cite{mueller2022instant}, and learned latent code conditioning. Among these, the INCODE architecture~\cite{kazerouni2023incode} demonstrated the most promising results in terms of reconstruction quality and convergence speed, while maintaining a low parameter count in our dynamic tomography setting.

\begin{figure*}[!h]
    \centering
    \includegraphics[width=1\linewidth]{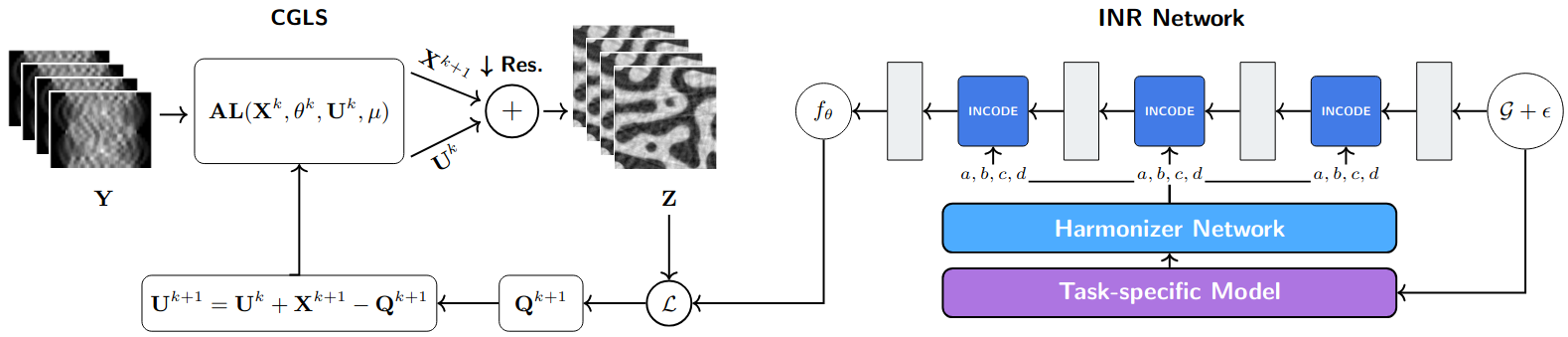}
    \caption[Network architecture]{\footnotesize{Network architecture. The reconstruction pipeline begins with an ADMM pass, where the input sinograms are used to compute an updated sequence of image estimates $\mathbf{X}$. These estimates are then combined with the Lagrange multipliers to form the intermediate variable $\mathbf{Z}$, which is subsequently downsampled. The INR optimization phase then begins. A coarse evaluation grid $\mathcal{G}$, perturbed with random offsets $\epsilon$, is used to sample the coordinates and evaluate the neural model's output. An illustration of the INR architecture is provided, explicitly incorporating the INCODE module. Standard neural network layers are shown in gray. At each evaluation step, the output image generated by the INR is compared to the downsampled version of $\mathbf{Z}$. The loss is computed from this comparison, along with spatial and temporal regularization terms. Once the model parameters are updated, the auxiliary variables $\mathbf{Q}$ , which represent the INR outputs and $\mathbf{U}$ dual variables, are also updated. This iterative cycle continues until convergence.}}
    \label{fig:network}
\end{figure*}

\subsection{ADMM-based optimization}

To accelerate convergence during training, we adopt an ADMM-based optimization strategy developed in \cite{ACCE}, reformulated for our dynamic setting. Instead of directly minimizing the reconstruction loss over the INR parameters $\theta$, we decouple the data fidelity term from the implicit representation using auxiliary variables. For a static scenario, the minimization scheme is as follows:
\begin{equation}
    \min_\theta \; \frac{1}{2} \| \mathbf{P}\left(\mathcal{T}(f_\theta(\mathcal{G}))\right) - \mathbf{y} \|_2^2,
\end{equation}
and is equivalently rewritten in a constrained form:
\begin{equation}
    \min_{\theta, \mathbf{x}} \; \frac{1}{2} \|  \mathbf{P}\mathbf{x}- \mathbf{y} \|_2^2 \quad \text{subject to} \quad \mathbf{x} = \mathcal{T}(f_\theta(\mathcal{G})),
\end{equation}
which gives the following rescaled augmented Lagrangian:
\begin{align}
    AL(\mathbf{x}, \theta, \mathbf{u}, \mu) =  & \frac{1}{2} \|  \mathbf{P}\mathbf{x} - \mathbf{y} \|_2^2 \notag \\ &+ \frac{\mu}{2} \|\mathbf{x} - \mathcal{T}\{f_\theta(\mathcal{G)}\} + \mathbf{u} \|_2^2.
\end{align}
where $\mathbf{u} \in \mathbb{R}^n$ has the interpretation as a vector of Lagrange multipliers.
We extend the original formulation to the dynamic case by solving a sequence of \(T\) such problems, one per time frame, while sharing the INR parameters across time. At each ADMM iteration \(k\), the updates are as follows:

\begin{itemize}
  \item \textbf{\(x\)-update (frame-wise):} for each \(t = 0, \dots, T{-}1\), we solve
\begin{align}
\mathbf{x}_t^{k+1} = \arg\min_{\mathbf{x}} \;
&\frac{1}{2} \| P_{\Theta_t} \mathbf{x} - \mathbf{y}_t \|_2^2 \notag \\
&+ \frac{\mu}{2} \| \mathbf{x} - \mathbf{q}_t^k + \mathbf{u}_t^k \|_2^2,
\label{eq:xt_update}
\end{align}
  
which is a regularized least-squares problem solved efficiently using conjugate gradient least square (CGLS).
  
  \item \textbf{\(\theta\)-update (shared across time):} we update the INR parameters by minimizing the mismatch with the current $x_t$ estimates and applying spatio-temporal regularization:
\begin{align}
\theta^{k+1} = \arg\min_\theta \;  \mathcal{L}\left(\mathbf{X}, \mathbf{Z} \right)  + \lambda_s \mathrm{TV}_\mathrm{space}(\mathbf{X}) \notag \\
+ \lambda_t \mathrm{TV}_\mathrm{time}(\mathbf{X}),
\label{eq:theta_update}
\end{align}
here \( \mathbf{X} = [\mathbf{x}_t]_{t=0}^{T-1} \) denotes the sequence of reconstructed images over time, where each frame \( \mathbf{x}_t \in \mathbb{R}^n \) is generated from the INR. 
Similarly, \( \mathbf{Z} = [\mathbf{x}_t^{k+1} + \mathbf{u}_t^k]_{t=0}^{T-1} \) combines the current estimate and dual variable for each frame, and is used as target in the parameter update step.

  \item \textbf{\(q\)-update:} for each \(t = 0, \dots, T{-}1\), we evaluate the current INR:
    \begin{equation}
        \mathbf{q}_t^{k+1} = \mathcal{T}(f_{\theta^{k+1}}(\mathcal{G}, t)).
    \end{equation}

  \item \textbf{\(u\)-update:} for each \(t = 0, \dots, T{-}1\), dual variables are updated via:
  \begin{equation}
  \mathbf{u}_t^{k+1} = \mathbf{u}_t^k + \mathbf{x}_t^{k+1} - \mathbf{q}_t^{k+1}.
\end{equation}
\end{itemize}

This alternating scheme allows us to decouple the projection-heavy \(x\) updates from INR training, significantly accelerating convergence. The CGLS algorithm requires access to an adjoint operator \( P^T \) that is mathematically consistent with the forward projection \( P \). However, in practice, many available implementations of the Radon forward and back-projection operators, such as those provided by \textit{scikit-image}~\cite{scikitimage} and \textit{torch-radon}~\cite{torch_radon}, do not satisfy the mathematical adjointness condition. As a result, the fundamental relationship given by:
\begin{equation}
    \langle P\mathbf{x}, \mathbf{y} \rangle = \langle \mathbf{x}, P^T \mathbf{y} \rangle
\end{equation}
may be violated numerically, compromising a core assumption that underpins the convergence guarantees of CGLS.

To address this, we implemented custom forward and adjoint operators using the \texttt{PyLops}~\cite{pylops} library and GPU-accelerated computations with \texttt{CuPy}, inspired by implementations commonly adopted in tomographic reconstruction literature~\cite{skimage1,skimage2}. To ensure memory-efficient interoperability between PyTorch and CuPy, we leverage the \texttt{DLPack} protocol, enabling zero-copy, in-place sharing of GPU tensors across frameworks.

\subsection{Improving memory efficiency during training}
\noindent We applied two practical architectures to ensure memory efficiency during model training. Following the jittered sampling approach of~\cite{4D,NERF}, we improve scalability by evaluating the INR on a lower-resolution spatial grid at each optimization step. To preserve the continuity of the implicit signal, random perturbations called jitters are applied to the coarse spatial coordinates. The CGLS iterations are still made at full resolution, allowing details preservation. 

To further address memory constraints during training, we apply the backward pass separately for each frame rather than accumulating gradients across the entire sequence. As a consequence, temporal regularization is computed in a pairwise fashion between consecutive frames $(t-1, t)$, which allows us to avoid retaining all intermediate frames within the computational graph. To mitigate directional bias and to enforce a more symmetric temporal consistency, we alternate the processing order of the sequence at each outer iteration, traversing it in chronological order for even iterations and in reverse (anti-chronological) order for odd ones. This simple strategy promotes bidirectional temporal smoothness without increasing memory requirements. This improves the average performance across frames while reducing variability, particularly for the first frame, which otherwise lacks temporal regularization.

Figure~\ref{fig:network} provides an overview of the ADMM-based optimization scheme and the INR architecture, while the pseudo-code is presented in Algorithm~\ref{alg:admm-dynamic}.

\begin{algorithm}[t]
\caption{ADMM-INR algorithm}
\label{alg:admm-dynamic}
\begin{algorithmic}[1]
\Statex \textbf{Input:} Temporal sinograms $\mathbf{Y} = \{ \mathbf{y}_t \}_{t=0}^{T-1}$

\State \textbf{Init:} $\theta^0$ (INR), $\mathbf{x}_t^0 \gets \mathbf{FBP}_{\Theta_t}(\mathbf{y_t})$, $\mathbf{u}_t^0 \gets \mathbf{0}$, $\mathbf{q}_t^0 \gets \mathcal{T}(f_{\theta^0}(\mathcal{G},t))$ for $t=0\ldots T{-}1$
 \vspace{2pt}
\For{$t=0$ \textbf{to} $T-1$}
   \State $
\begin{aligned}
\mathbf{x}_t^{k+1} \gets  &  \arg\min_{\mathbf{x}} \;
\frac{1}{2} \| P_{\Theta_t} \mathbf{x} - \mathbf{y}_t \|_2^2 \notag \\ 
&+ \frac{\mu}{2} \| \mathbf{x} - \mathbf{q}_t^{k} + \mathbf{u}_t^k \|_2^2
\end{aligned}
$ \Comment{CGLS}
  \EndFor
  \vspace{2pt}
  \State \textbf{Set frame order} $\mathcal{O}_k$ \Comment{chronological if $k$ even, reverse if $k$ odd} 
  \State $\bar{\mathbf{p}}_{t-1} \gets \varnothing$ \Comment{previous prediction}
  \For{$t \in \mathcal{O}_k$}
      
  \State $\tilde{\mathbf{z}}_t \gets \mathcal{D}_s\!\left(\mathbf{x}_t^{k+1} + \mathbf{u}_t^{k}\right)$ \hfill \Comment{downsampled target}
  \State $\mathbf{p}_t \gets \mathcal{T}\!\big(f_\theta(\tilde{\mathcal{G}}, t)\big)$ \hfill \Comment{INR output (low-res)}
  \State $\mathcal{L} \gets \|(\mathbf{p}_t - \tilde{\mathbf{z}}_t)\|_2^2$ \hfill \Comment{MSE}
  \If{$k > k_s$}
    \State $\mathcal{L} \gets \mathcal{L} + \lambda_s \,\|\nabla \mathbf{p}_t\|_1$ \hfill \Comment{spatial TV}
  \EndIf
  \If{$\bar{\mathbf{p}}_{t-1}\neq\varnothing$ \textbf{and} $k>k_t$}
    \State $\mathcal{L} \gets \mathcal{L} + \lambda_t \,\|\mathbf{p}_t - \bar{\mathbf{p}}_{t-1}\|_1$ \hfill \Comment{temporal TV}
  \EndIf 
  \State \textbf{backward}$(\mathcal{L})$
  \State $\bar{\mathbf{p}}_{t-1} \gets \mathrm{detach}(\mathbf{p}_t)$ 
\EndFor
  \vspace{2pt}
  \For{$t=0$ \textbf{to} $T-1$}
    \State $\mathbf{q}_t^{k+1} \gets \mathcal{T}(f_{\theta^{k+1}}(\mathcal{G},t))$
  \EndFor
  \vspace{2pt}
  \For{$t=0$ \textbf{to} $T-1$}
    \State $\mathbf{u}_t^{k+1} \gets \mathbf{u}_t^{k} + \mathbf{x}_t^{k+1} - \mathbf{q}_t^{k+1}$
  \EndFor
\State \Return $\theta^{K}$
\end{algorithmic}
\end{algorithm}

\section{Experiments}
\subsection{Simulated X-ray computed tomography data set}

To evaluate the proposed framework under controlled yet physically plausible dynamics, we simulate a time-evolving microstructural process governed by spinodal decomposition \cite{spinodal-paper}. This phenomenon describes phase separation in binary mixtures and is modeled by the Cahn–Hilliard equation \cite{CH}:

\begin{equation}
\frac{\partial c(\mathbf{r}, t)}{\partial t} = \nabla \cdot \left[ M(c) \nabla \left( \frac{\partial f}{\partial c} - \epsilon \nabla^2 c \right) \right],
\label{eq:CH}
\end{equation}
here, $c(\mathbf{r}, t)$ denotes the concentration field, $f(c)$ is the chemical free energy density, $M(c)$ the mobility, and $\epsilon$ a gradient energy coefficient controlling interface width.

Spinodal decomposition offers a rich yet controlled and reproducible spatio-temporal dynamic, characterized by evolving interfaces and microstructural complexity. It has been used in prior work to test dynamic CT frameworks \cite{TIMBR}.

Beyond serving as a synthetic benchmark, spinodal decomposition constitutes a fundamental mechanism in materials science, enabling the design of architected materials with tailored properties such as high energy absorption \cite{spino-highenergy} or extreme mechanical resilience \cite{spino-mechanical}. It has also been leveraged to engineer battery electrodes with enhanced capacity and long term cycling stability \cite{batteries-spino, batteries-spino2}. Its influence even extends to nuclear materials, where it modulates the thermomechanical properties of metallic fuels under irradiation \cite{YAO_SPINODAL_1, YAO_SPINODAL_2}.

\begin{figure}[h!]
    \centering
    \includegraphics[width=1\linewidth]{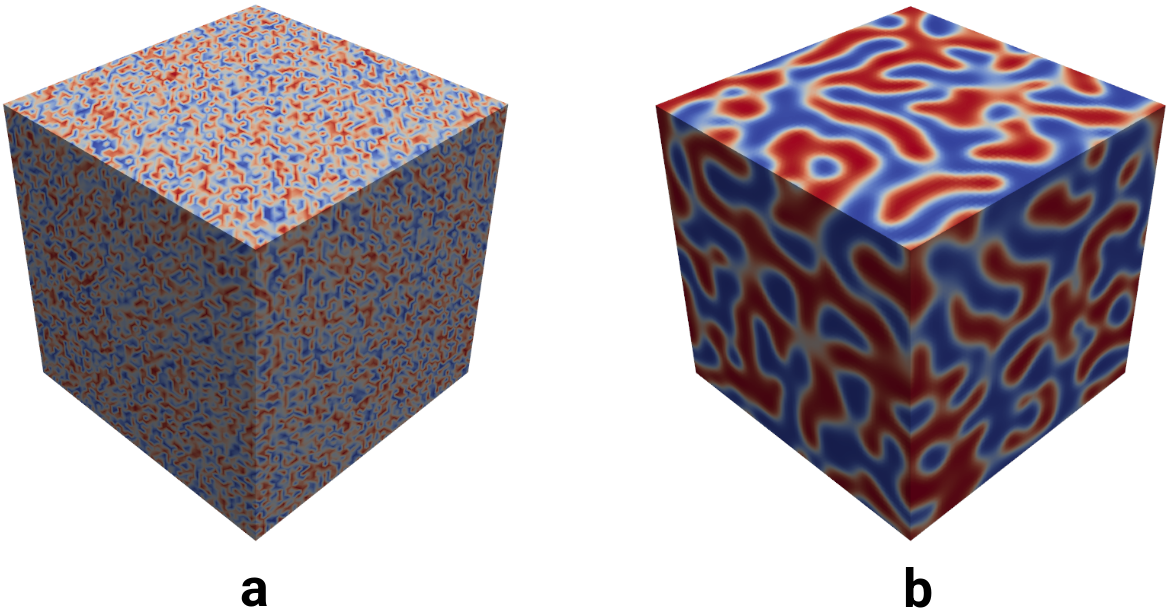}
    \caption{3D Phase-field simulation of spinodal decomposition performed using a semi-implicit spectral method. (a) Initial , (b)  final microstructure respectively. The simulation domain consists of a $64 \times 64 \times64$ grid, with a constant mobility  M of 1, and an initial homogeneous composition c of 0.5.}
    \label{fig:3D_phase_field}
\end{figure}

For the 2D case, we use the open-source solver Prism-pf, which employs a matrix-free finite element method for efficient numerical computation~\cite{PRISM-PF}. This produces a sequence of N high-resolution frames representing the evolving object over time. Meanwhile, for 3D data generation (Figure~\ref{fig:3D_phase_field}), we used a Python-based implementation using a semi-implicit spectral method~\cite{complas}, as the use of Prism-pf proved computationally and memory intensive, making data processing impractical.

To emulate dynamic XCT acquisition with motion, the sequence of images is divided into \(T\) temporal groups of \(P = N/T\) frames, each assigned a unique set of projection angles \(\Theta_t = \{\theta_1^{(t)}, \dots, \theta_P^{(t)}\} \subset \Theta\) following the interlaced acquisition scheme. In each group, the \(i\)-th frame is projected at angle \(\theta_i^{(t)}\), producing a sinogram composed of projections of different object states. As a representative case to assign realistic X‑ray attenuation, we use the Al–Cu alloy system (matrix $\alpha$‑Al and Al$_2$Cu precipitate). The voxel resolution is 3$\times$3$\times$3 $\mu$m$^3$, and at 60 keV the two phases have attenuation coefficients of 0.0750 mm$^{-1}$ and 0.4303 mm$^{-1}$, respectively \cite{hubbell2004xray}. For visualization and evaluation, we define the ground truth at time \(t\) as the filtered back projection (FBP) of the central frame of group \(t\), approximating the average object state.

\subsection{Model evaluation metrics}

We evaluate reconstruction quality using two standard metrics: peak signal-to-noise ratio (PSNR) and structural similarity index measure (SSIM). PSNR quantifies pixel-wise fidelity, while SSIM captures perceptual similarity.

PSNR is defined as:
\begin{equation}
    \text{PSNR}(x, x^*) = 10 \cdot \log_{10} \left( \frac{\text{MAX}^2}{\text{MSE}(x, x^*)} \right),
\end{equation}
where $x$ is the reconstruction, $x^*$ the ground truth, and $\text{MAX}$ the maximum pixel value. The mean squared error (MSE) is given by:
\begin{equation}
    \text{MSE}(x, x^*) = \frac{1}{N} \sum_{i=1}^{N} (x_i - x^*_i)^2.
\end{equation}

SSIM~\cite{SSIM} evaluates structural similarity over local patches:
\begin{equation}
    \text{SSIM}(x, x^*) = \frac{(2 \mu_x \mu_{x^*} + C_1)(2 \sigma_{x x^*} + C_2)}{(\mu_x^2 + \mu_{x^*}^2 + C_1)(\sigma_x^2 + \sigma_{x^*}^2 + C_2)},
\end{equation}
where $\mu$, $\sigma^2$ and $\sigma_{x x^*}$ denote local means, variances, and covariance; $C_1$, $C_2$ are small constants to stabilize the expression.

Together, PSNR and SSIM provide a complementary assessment of reconstruction quality, from both pixel-level accuracy and perceptual consistency.

\subsection{Experimental setup}
All experiments were performed using PyTorch on an NVIDIA GeForce RTX 4060 GPU with 8GB of memory. We employ the Adam optimizer \cite{ADAM} with a learning rate scheduler that decreases the learning rate progressively to support convergence. Each ADMM iteration consisted of 20 CGLS steps followed by 50 INR updates. Unless otherwise stated, the INR was implemented as a three-layer multilayer perceptron (MLP) with 256 hidden units per layer, using Gaussian positional encoding with a scale factor of 5 and a mapping input of 256. The INCODE harmonizer was truncated at the fifth order and employed the SiLU activation function. A pretrained ResNet34 from the PyTorch model zoo served as the encoder. 

For TIMBIR reconstructions, the parameters of the q-generalized Gaussian Markov random field (qGGMRF) prior were tuned to optimize reconstruction quality. In our case, we adjusted the spatio-temporal regularization hyperparameters, with all configurations detailed in the accompanying code (sec. \ref{sec:code}). As our method is iterative, we track the mean residual $\mathbf{x} -\mathbf{q}$ across temporal frames at each iteration and retain the model weights corresponding to the iteration that minimizes this residual.

\section{Results and discussion}
\subsection{First experiment}
\label{sec:first}
\noindent Let $\delta t$ denote the acquisition time per projection. We consider three configurations that span the same total acquisition duration $T = N_{\theta} \cdot \delta t$, corresponding to the full temporal evolution of the dynamic process. Specifically, we set $(N_{\theta}, \delta t)$ to $(256, \delta t)$, $(128, 2\delta t)$, and $(64, 4\delta t)$, respectively, with $K = 16$ fixed in all cases. By increasing the acquisition time per projection in the sparser settings, we maintain the same temporal range from the initial to the final state across configurations. This allows us to assess reconstruction robustness under decreasing angular sampling density, while keeping the dynamic acquisition window fixed. For consistency with TIMBIR, which, to our understanding, is constrained to circular reconstruction domains, we restrict the number of detectors to the image width. 

\begin{figure}[h!]
    \centering
    \includegraphics[width=0.95\linewidth]{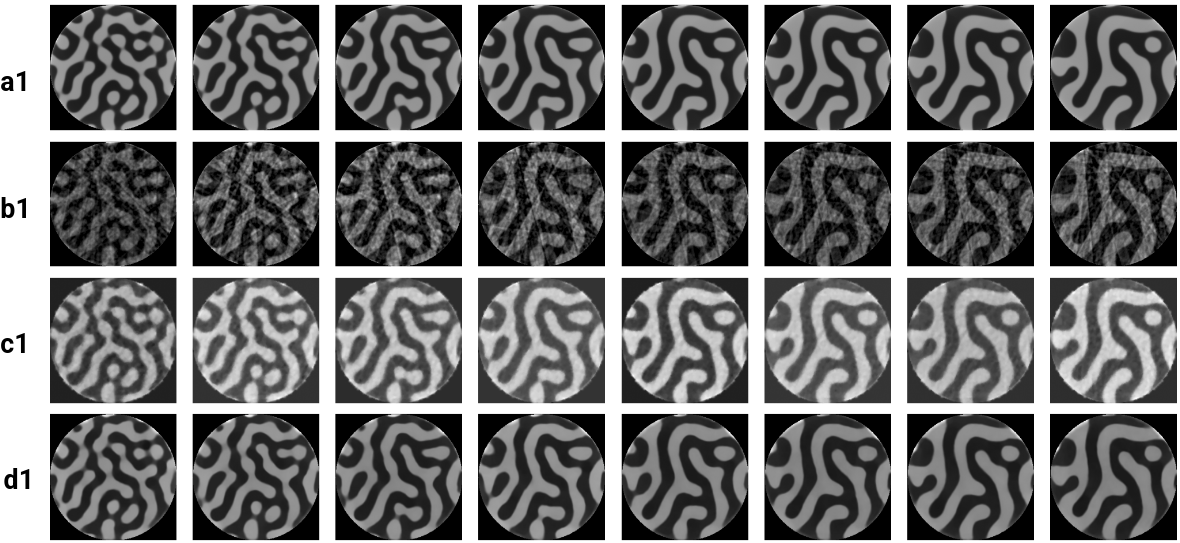}
    \vspace{0.5cm}
    
    \includegraphics[width=0.95\linewidth]{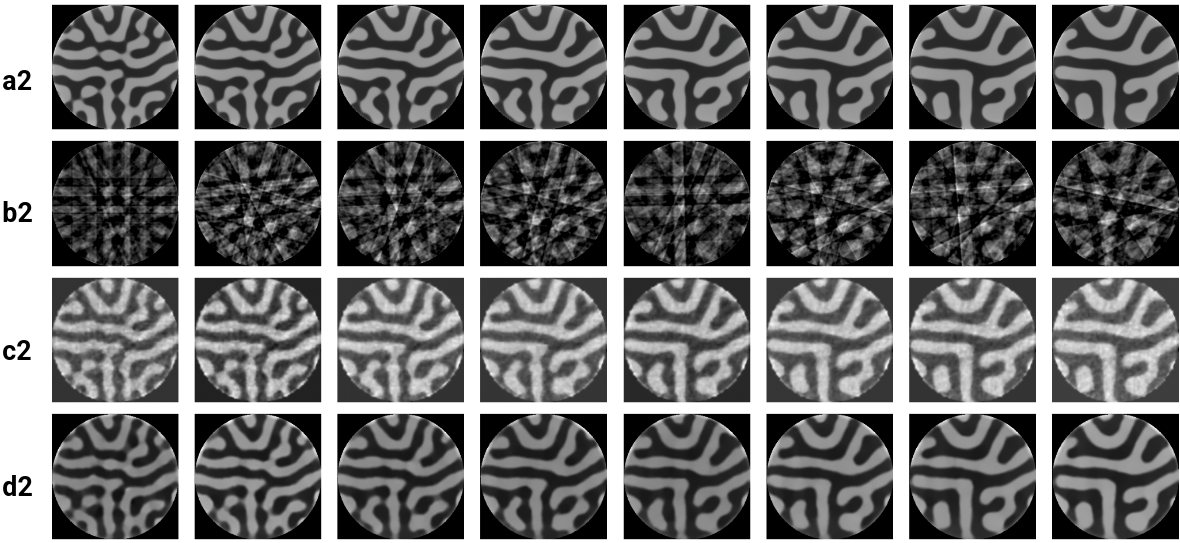}
    \vspace{0.5cm}
    
    \includegraphics[width=0.96\linewidth]{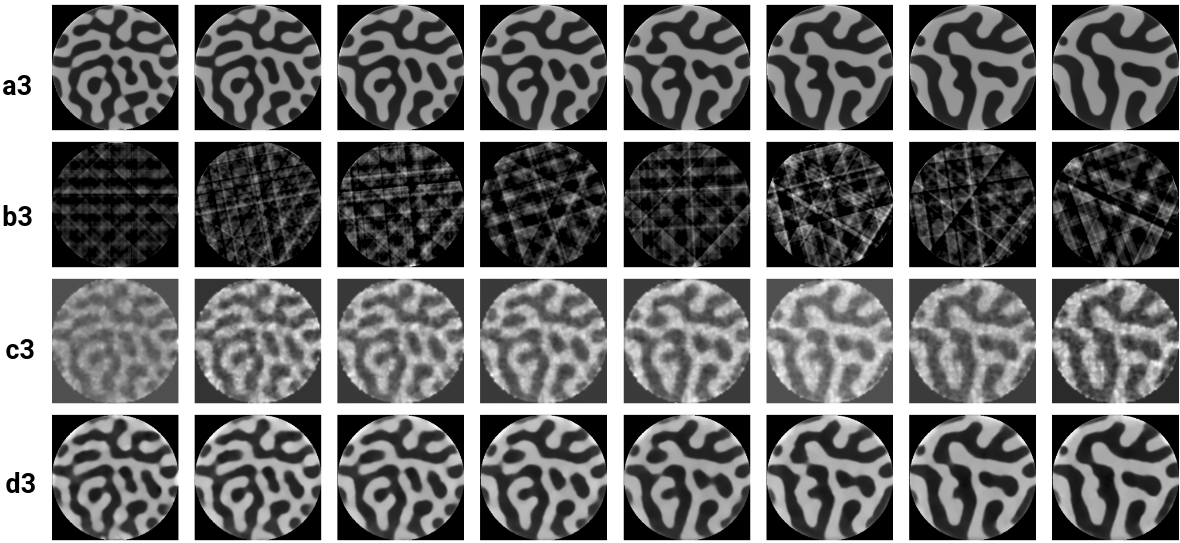}
   \caption[Reconstruction results under varying acquisition settings]{Reconstruction results under varying acquisition settings. \textbf{1} | Full sampling with $N_{\theta} = 256$, \textbf{2} | Moderate undersampling with $N_{\theta} = 128$, \textbf{3} | Severe undersampling with $N_{\theta} = 64$. (a) Ground truth, (b) FBP reconstruction, (c) TIMBIR reconstruction, (d) Our method. Although $K = 16$ time frames were reconstructed, only every second frame is shown to reduce visual clutter.}
    \label{fig:configs}
\end{figure}

Figure \ref{fig:configs} and Table~\ref{tab:ncc_psnr_first} reports reconstruction performance across acquisition settings with decreasing angular sampling density. As the number of projections is reduced from 256 to 64, both methods show a drop in performance, reflecting the increased difficulty of the inverse problem under sparse angular coverage. For TIMBIR, PSNR decreases from 16.57dB to 14.14dB, and SSIM from 0.723 to 0.512. In contrast, our method shows a smaller decline, from 24.83dB to 22.86dB in PSNR, and from 0.903 to 0.861 in SSIM.

Across all configurations, our approach maintains a consistent advantage of approximately 7–9dB in PSNR and 0.15–0.35 in SSIM over TIMBIR. While both methods degrade as angular sampling becomes sparser, the relative stability of our model suggests that the implicit representation is able to better exploit spatio-temporal continuity in the data. These results highlight the potential of INR-based reconstruction strategies to maintain quality even under challenging acquisition conditions.

\begin{table*}[ht]
    \centering
    \footnotesize
    \caption{
        Reconstruction performance across acquisition configurations. Bold indicates best performance.
    }
    \setlength{\tabcolsep}{4pt}
    \renewcommand{\arraystretch}{1.2}
    \begin{tabular}{c| *2{c} | *2{c} | *2{c} }   
        \toprule
        & \multicolumn{2}{c|}{$(256,\ \delta t)$}
        & \multicolumn{2}{c|}{$(128,\ 2\delta t)$}
        & \multicolumn{2}{c}{$(64,\ 4\delta t)$} \\
        \cline{2-7}
        & PSNR &  SSIM 
        & PSNR  & SSIM
        & PSNR & SSIM\\
        \hline
        TIMBIR 
        & 16.57 $\pm$ 2.30 & 0.723 $\pm$ 0.029 
        & 16.54 $\pm$ 1.45 & 0.669 $\pm$ 0.027 
        & 14.14 $\pm$ 1.08 & 0.512 $\pm$ 0.020 \\
        Ours   
        & \textbf{24.83 $\pm$ 0.60} & \textbf{0.903 $\pm$ 0.008} 
        & \textbf{23.23 $\pm$ 0.60} & \textbf{0.881 $\pm$ 0.014}
        & \textbf{22.86 $\pm$ 0.85} & \textbf{0.861 $\pm$ 0.025} \\
        \bottomrule
    \end{tabular}
    \label{tab:ncc_psnr_first}
\end{table*}

\subsection{Impact of microstructural complexity}

\noindent Understanding the behavior of the model under varying microstructural complexity is essential for robust performance analysis. While the previous results provided promising insights, they did not explicitly account for the intrinsic spatial complexity of the input data. In this section, we investigate how increasing microstructural complexity affects model behavior.

Previous simulations were generated on a high-resolution mesh and subsequently mapped onto images. In earlier experiments, only a cropped 256$\times$256 region of these high-resolution images was used as input, thereby limiting the spatial variability and fine-scale details present in the data. In the present case, while the model input size remains fixed at 256$\times$256, the cropped regions are now extracted from larger areas of 512$\times$512 and 1024$\times$1024, respectively. This increase in crop size enhances the structural complexity and information present in the input, allowing us to evaluate the sensitivity of the model to more complex and diverse textures and longer-range spatial dependencies.

Quantifying image complexity, however, is not a straightforward task. Classical information-theoretic measures such as Shannon entropy are ill-suited because entropy is calculated without considering spatial structures~\cite{Complexity}. To address this, we rely on the concept of spatial information (SI)~\cite{Complexity}. SI is based on the magnitude of local spatial gradients and captures the amount of local structural variation present in an image.  Let $s_h$ and $s_v$ denote gray-scale images filtered with horizontal and vertical Sobel kernels, respectively. Given an image of P pixels, the mean magnitude of SI at every pixel is given by:
\begin{equation}
    SI_r = \sqrt{s_h^2 + s_v^2}, \quad 
    SI_{\text{mean}} = \frac{1}{P} \sum SI_r.
\end{equation}
To ensure contrast invariance, SI is computed on a binarized version of the image. For temporal sequences, we retain the maximum SI\textsubscript{mean} across all frames.  For these cases we increase the Gaussian positional encoding to 10 and 15 with a mapping input size of 512.

\begin{figure}[t]
    \centering
    \includegraphics[width=1\linewidth]{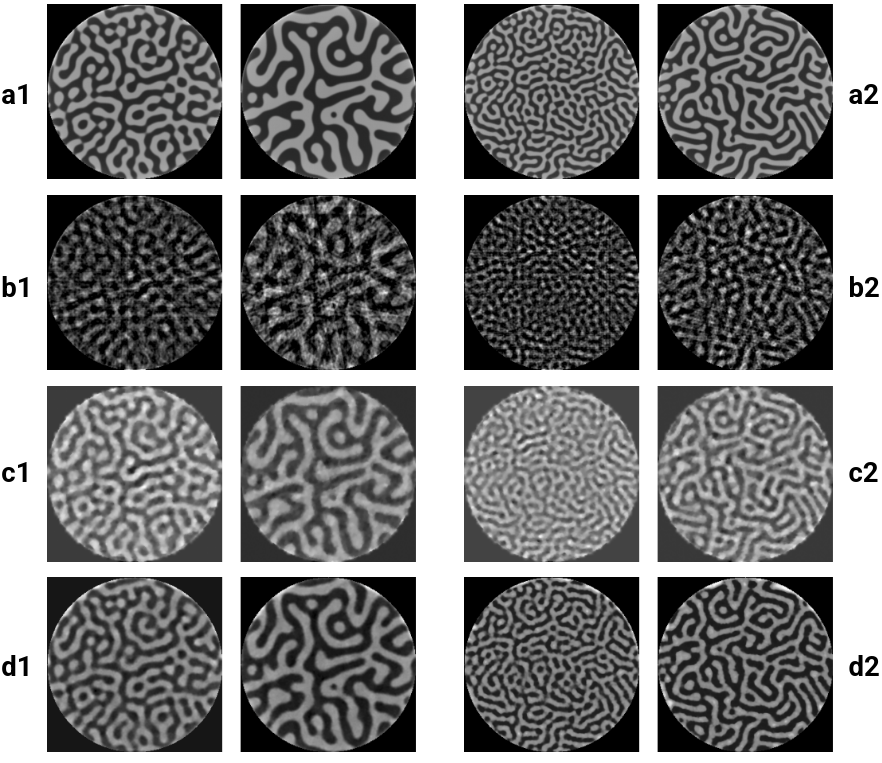}
    \caption[Impact of spatial complexity on reconstruction performance]{Impact of spatial complexity on reconstruction performance. \textbf{1} | Case 1, \textbf{2} | Case 2. (a) Ground truth, (b) FBP reconstruction, (c) TIMBIR reconstruction, (d) Our method. Although $K = 16$ time intervals were reconstructed, we only show the first and last ones to reduce visual clutter.}
    \label{fig:complexity}
\end{figure}

Figure~\ref{fig:complexity} and Table~\ref{tab:ncc_psnr_complexity} reports reconstruction performance across two levels of spatial complexity, measured using the Spatial Information (SI) metric. When the spatial content is less complex (SI = 0.724), both methods perform better overall, but our approach shows a notable improvement over TIMBIR, with a PSNR gain exceeding 3 dB and a SSIM increase of nearly 0.06. In the more challenging setting (SI = 1.056), performance drops for both methods, as expected, due to the increased structural variability. Nevertheless, our method maintains a consistent advantage in both PSNR and SSIM.

Importantly, our approach also exhibits reduced temporal variability, as evidenced by the lower standard deviation values, particularly in the high-complexity case. This suggests that the method not only achieves higher average fidelity, but does so more consistently across the temporal sequence. Overall, these results indicate that the method remains robust across varying levels of spatial detail, while maintaining temporal stability.

\begin{table*}[ht]
    \centering
    \footnotesize
    \caption{
        Reconstruction performance for varying crop sizes and spatial complexity.  SI = Spatial Information. Bold indicates best result per metric.
    }
    \setlength{\tabcolsep}{12pt}
    \renewcommand{\arraystretch}{1.2}
    \begin{tabular}{c| *2{c} | *2{c} }   
        \toprule
        & \multicolumn{2}{c|}{SI = 0.724}
        & \multicolumn{2}{c}{SI = 1.056} \\
        \cline{2-5}
        & PSNR & SSIM 
        & PSNR & SSIM \\
        \hline
        TIMBIR 
        & 17.45$\pm$ 1.27 & 0.729 $\pm$ 0.015
        & 16.33  $\pm$ 1.47 & 0.730 $\pm$ 0.038 \\
        Ours
        & \textbf{20.78 $\pm$ 0.30} & \textbf{0.786 $\pm$ 0.009}
        & \textbf{17.98 $\pm$ 0.22} & \textbf{0.742 $\pm$ 0.011} \\
        \bottomrule
    \end{tabular}
    \label{tab:ncc_psnr_complexity}
\end{table*}

\subsection{Noise resilience}
\noindent While the previous XCT scan model assumes ideal, noise-free measurements, real-world acquisitions are inherently affected by noise, particularly under low-dose conditions, where photon counts are limited and detector imperfections may be present. To evaluate the robustness of our model under realistic noise conditions, we simulate X-ray acquisition noise using a Poisson model consistent with the stochastic nature of photon detection. Given a noise-free sinogram value \( p \), the expected photon count along a ray is modeled as \( \mu = \text{dose} \cdot e^{-p} \), and noisy measurements are drawn from \( I \sim \mathrm{Poisson}(\mu) \). The corresponding noisy sinogram is then obtained by:
\begin{equation}
    p_{\text{noisy}} = \log\left( \frac{\text{dose}}{I} \right).
\end{equation}
Noise severity is controlled via the dose parameter, with lower doses inducing a low signal-to-noise ratio. We evaluate three noise levels using dose values of $20 \times 10^3$, $5 \times 10^3$, and $1 \times 10^3$, applied under the first acquisition configuration (Section~\ref{sec:first}). This setup enables us to evaluate whether the inductive bias of INRs alone is sufficient to ensure robustness against noise. Additionally, we consider a second formulation based on a weighted least squares (WLS) model, which explicitly incorporates the statistical nature of Poisson noise. In this setting, the data fidelity term becomes:
\begin{equation}
    \frac{1}{2}\| W^{1/2} (\mathbf{P} \mathbf{x} - \mathbf{y}) \|_2^2,
\end{equation}
where \( W = \mathrm{diag}(w) \) is a diagonal matrix with weights set proportional to the measured photon counts, i.e., \( w \propto I \). This choice approximates the inverse noise variance under a Poisson model and better reflects the heteroscedastic nature of the data, thereby improving robustness in low-dose settings where the noise level varies significantly across detectors.

\begin{figure}[!htbp]
    \centering
    \includegraphics[width=1\linewidth]{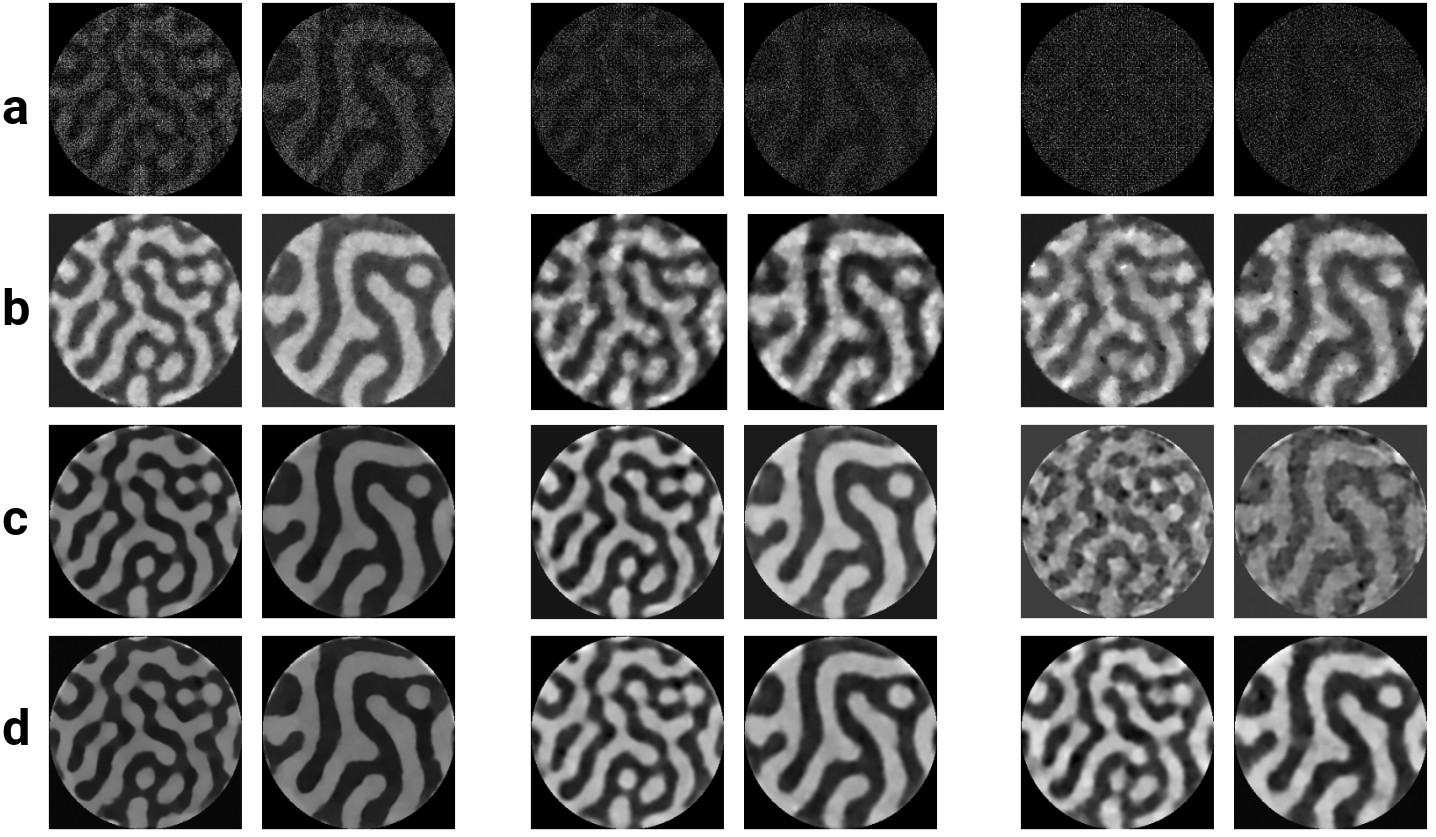}
    \caption[Reconstruction for different levels of noise]{Reconstruction of configuration 1 for different levels of dose. From left to right, dose = \{20, 5, 1\}$\times$10$^3$. (a) FBP reconstruction, (b) TIMBIR reconstruction, (c) Our method, (d) Ours* = Ours with WLS. Although $K = 16$ time intervals were reconstructed, we only show the first and last reconstructions to reduce visual clutter.}
    \label{fig:noise}

\end{figure}

Figure~\ref{fig:noise} and Table~\ref{tab:noise} present the corresponding reconstruction results. As shown in Table~\ref{tab:noise}, our original formulation (\textit{Ours}) performs best at the highest dose level, achieving the highest SSIM (0.873) and PSNR (21.49~dB). This suggests that the inherent inductive bias of the INR is sufficient to ensure high-quality reconstruction in low-noise conditions.

However, as the dose decreases, noise becomes more prominent and reconstruction becomes more challenging. In these regimes, the WLS-based formulation (\textit{Ours$^*$}), which introduces a data fidelity term weighted according to the estimated noise variance, demonstrates significantly improved robustness. At a dose of \(5 \times 10^3\), \textit{Ours$^*$} improves the SSIM from 0.778 (\textit{Ours}) to 0.799, while also yielding the best PSNR (21.26~dB). The benefit becomes even more pronounced in the most challenging scenario of \(1 \times 10^3\), where \textit{Ours$^*$} achieves a SSIM of 0.713, substantially outperforming both the original formulation (0.559) and TIMBIR (0.556), despite TIMBIR achieving a slightly higher PSNR (17.07~dB vs. 16.81~dB). Interestingly, in these experience, TIMBIR maintain the lowest variability for PSNR results.

\begin{table*}[ht]
    \centering
    \footnotesize
    \caption{
        Reconstruction performance at various dose levels (10$^3$). Ours$^*$ = Ours with WLS. Bold indicates best performance.
    }
    \setlength{\tabcolsep}{4pt}
    \renewcommand{\arraystretch}{1.2}
    \begin{tabular}{l| *2{c} | *2{c} | *2{c} }   
        \toprule
        & \multicolumn{2}{c|}{Dose = 20}
        & \multicolumn{2}{c|}{Dose = 5}
        & \multicolumn{2}{c}{Dose = 1} \\
        \cline{2-7}
        & PSNR & SSIM
        & PSNR & SSIM 
        & PSNR & SSIM \\
        \hline
        TIMBIR 
        & 17.57 $\pm$ 0.78 & 0.752 $\pm$ 0.012
        & 18.39 $\pm$ 0.45 & 0.587 $\pm$ 0.013
        & \textbf{17.067$\pm$ 0.612} & 0.556 $\pm$ 0.022 \\
        Ours
        &  \textbf{21.49 $\pm$ 0.86} & \textbf{0.873 $\pm$ 0.011}
        & 19.07 $\pm$ 1.53 & 0.778 $\pm$ 0.021
        & 16.68 $\pm$ 1.605 & 0.559 $\pm$ 0.029 \\
        Ours$^*$  
        &21.45 $\pm$ 1.39 & 0.865 $\pm$ 0.012
        & \textbf{21.26 $\pm$ 1.07} & \textbf{0.799 $\pm$ 0.012}
        & 16.81 $\pm$ 1.08 & \textbf{0.713 $\pm$ 0.022} \\
        \bottomrule
    \end{tabular}
    \label{tab:noise}
\end{table*}

\section{Toward a complete framework}
\subsection{Detector non-idealities}

\begin{figure}
    \centering
    \includegraphics[width=1\linewidth]{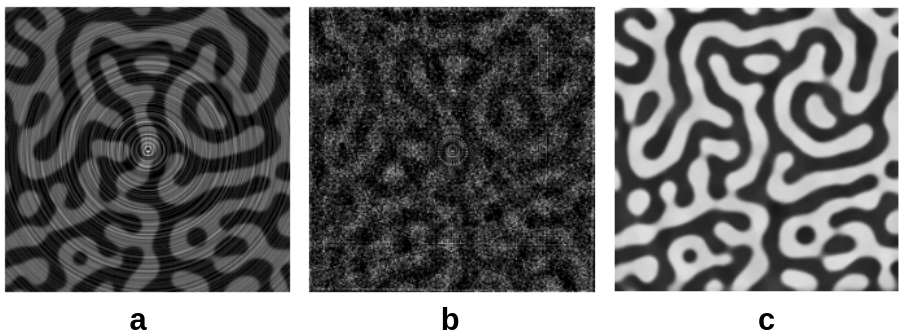}
    \caption{Ring artifacts. (a) FBP reconstruction from a full-view ($N_{\theta} = 256$) acquisition on a static image, illustrating the strength of the injected ring artifacts. This provides a visual benchmark for detector-induced bias. (b) FBP reconstruction of the first individual subframe in a temporally interlaced setting with $K = 16$ subframes, where the reduced angular density makes artifacts less apparent. (c) Reconstruction using our method under the same acquisition setup. A detector count of $N_d = 363$ ensures complete coverage.}
    \label{fig:rings}
\end{figure}

Systematic detector imperfections can give rise to structured artifacts in computed tomography, most notably concentric rings in the reconstructed volume. These ring artifacts (Figure~\ref{fig:rings}.a) originate from persistent inconsistencies across detector bins—such as gain variations or electronic drift, and manifest in the sinogram domain as additive biases that are invariant across projection angles but vary along the detector axis~\cite{rivers1998tutorial}.

To model these effects, we introduce a vector $\mathbf{c} \in \mathbb{R}^{N_d}$ representing a static, angle-invariant additive bias for each detector bin. At each time step $t$, a linear operator $C_t$ replicates $\mathbf{c}$ across the angular dimension, yielding the following forward model:
\begin{equation}
    \mathbf{y}_t = P_{\Theta_t} \mathbf{x}_t + C_t \mathbf{c},
\end{equation}
where $\mathbf{y}_t$ is the measured sinogram and $P_{\Theta_t} \mathbf{x}_t$ is the ideal projection at time $t$.

This model is incorporated into our ADMM-based optimization by treating $\mathbf{c}$ as an auxiliary variable. At each iteration, after updating all $\mathbf{x}_t$, we compute the sinogram residuals:
\begin{equation}
    \mathbf{r}_t = \mathbf{y}_t - P_{\Theta_t} \mathbf{x}_t^{(k+1)},
\end{equation}
and stack them across time and angles to form a residual matrix $R \in \mathbb{R}^{M \times N_d}$, where $M = \sum_{t} |\Theta_t|$ is the total number of rays. Each column of $R$ aggregates the residuals corresponding to a single detector bin over all angles and time frames.

To robustly estimate $\mathbf{c}$ from $R$, we solve the following weighted least-squares problem using Iteratively Reweighted Least Squares (IRLS) with a Huber loss \cite{Huber}:
\begin{equation}
    \mathbf{c}^{(k+1)} = \arg\min_{\mathbf{c}} \sum_{m=1}^{M} \sum_{d=1}^{N_d} \rho_\delta(R_{m,d} - c_d),
\end{equation}
where $\rho_\delta$ is the standard Huber function with threshold $\delta$. The IRLS procedure assigns lower weights to outliers, improving robustness~\cite{IRLS-1, IRLS-2, IRLS-3}. We enforce a zero-mean constraint on $\mathbf{c}^{(k+1)}$ to ensure identifiability and optionally apply one-dimensional smoothing (e.g., Tikhonov regularization) along the detector axis to promote spatial coherence. Alternatively, sparsity-promoting constraints could also be applied to $\mathbf{c}$~\cite{sparsity-1, sparsity-2}.

Delegating artifact reduction directly to the INR, for example, through an unsupervised multi-parameter inverse solving strategy~\cite{Wu2024UnsupervisedMI}, would be an appealing alternative. However, due to the variable splitting introduced by the ADMM framework, the projection operator is no longer explicitly available within the INR module.

The $\mathbf{x}_t$-update step in our ADMM framework solves the following regularized least-squares problem:
\begin{align}
    \mathbf{x}_t^{k+1} = \arg\min_{\mathbf{x}} \; 
    \frac{1}{2} & \left\| P_{\Theta_t} \mathbf{x} - (\mathbf{y}_t - C_t \mathbf{c}^{k}) \right\|_2^2 \notag \\
    &+ \frac{\mu}{2} \left\| \mathbf{x} - \mathbf{q}_t^k + \mathbf{u}_t^k \right\|_2^2.
\end{align}

While we retain the standard quadratic fidelity term in this work, it may be replaced by a robust alternative such as the Huber loss \cite{Huber, TIMBR} to ensure comprehensive treatment of zingers. This modification remains fully compatible with our ADMM-INR framework.

\subsection{Four-dimensional extension}
\label{sec:4D}
The previous sections and results have focused primarily on reconstructing 2D+t volumes. However, this approach can be extended to full four-dimensional (4D) volumetric reconstructions by incorporating the axial coordinate \(z\) into the INR, which now takes spatio-temporal coordinates \((x, y, z, t)\) as input. To enforce spatial consistency along the axial direction, the loss function is augmented with an additional axial continuity term, and the optimization loop includes an iteration over \(z\).

Nevertheless, training a single INR over the entire 3D+t volume is not computationally efficient. First, axial continuity provides meaningful regularization only within a limited spatial range. Second, optimizing a monolithic INR restricts parallelism and requires substantial computational resources on large datasets. We therefore adopt a more scalable strategy based on axial batching. Instead of reconstructing the full 4D volume at once, we divide it into smaller subproblems, each associated with a batch of \(Z_b\) contiguous axial slices. This allows us to leverage local axial context while maintaining efficient parallelization. Each subvolume, defined over \((x, y, z_k, t)\), is reconstructed independently by a dedicated INR.

\noindent Given a batch of temporal sinogram measurements 
\begin{equation}
    \mathbf{Y} = [\mathbf{Y}_0, \dots, \mathbf{Y}_{T-1}], \quad \text{with} \quad \mathbf{Y}_t \in \mathbb{R}^{N_{d} \times N_{\theta} \times Z_b},
\end{equation}
where \(T\) is the number of temporal acquisitions, \(N_d\) the number of detector elements, \(N_{\theta}\) the number of projection angles, and \(Z_b\) the number of jointly reconstructed axial slices. The final optimization problem for each axial subvolume becomes:
\begin{align}
\min_\theta \; & \mathcal{L}\left(P_{\boldsymbol{\Theta}}\{\mathbf{X}\}, \mathbf{Y} \right) + \lambda_s \,  \mathrm{TV}_\mathrm{space} 
 (\mathbf{X}) \notag \\ 
 &
 + \lambda_a \, \mathrm{TV}_\mathrm{axial}(\mathbf{X}) + \lambda_t \, \mathrm{TV}_\mathrm{time}(\mathbf{X}),
\end{align}
where \(P_{\boldsymbol{\Theta}}\) the forward projector is applied to the reconstructed volume \(\mathbf{X}\), parameterized by the INR weights \(\boldsymbol{\theta}\), and the TV terms promote smoothness across the spatial, axial, and temporal dimensions, respectively.

In this work, each batch is treated independently. However, alternative strategies may improve consistency across the full volume. For example, one may adopt a sliding window approach where neighboring axial subvolumes overlap, and impose coherence through additional coupling terms. A simple regularization between overlapping regions \(\Omega = \mathcal{W}_m \cap \mathcal{W}_n\) could be written as
\begin{equation}
    \lambda_c \sum_{z \in \Omega} \left\| x^{(m)}_z - x^{(n)}_z \right\|^2,
\end{equation}
or, in a stricter consensus setting, by enforcing equality constraints of the form
\begin{equation}
    x^{(m)}_z = x^{(n)}_z, \quad \forall z \in \Omega.
\end{equation}

A practical limitation of our current architecture is that INCODE cannot directly handle 4D input (\(N_d \times N_\theta \times Z_b \times T\)). One option is to replace the pretrained 2D feature extractor with a 3D convolutional architecture (e.g., ResNet3D-18), at the cost of increased complexity. Alternatively, the input can be flattened so that each slice is processed independently using the original 2D network. We adopt this lightweight strategy for simplicity and memory efficiency.

\begin{figure}[ht]
    \centering
    \includegraphics[width=1\linewidth]{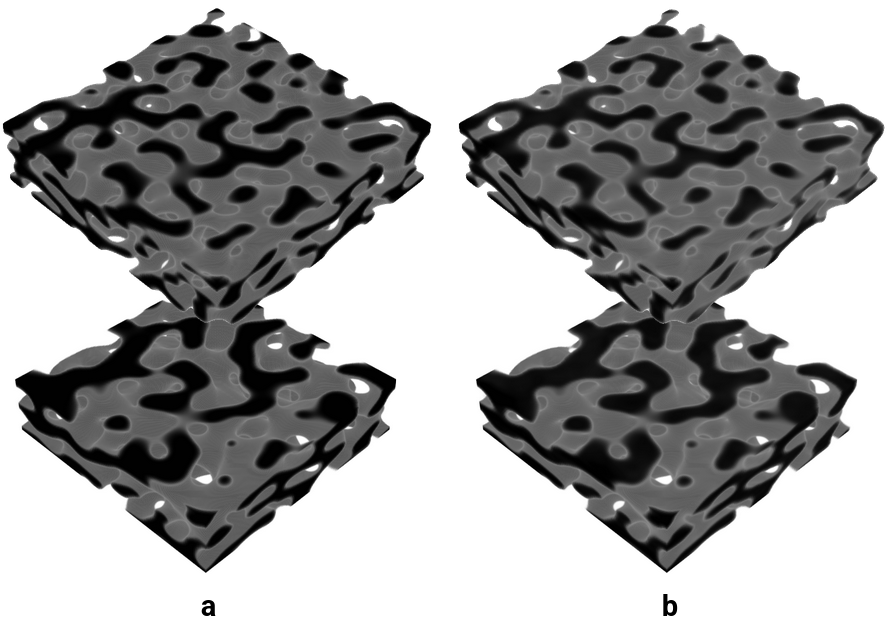}
    \caption[3DT reconstruction results]{3DT reconstruction results of a $256 \times 256 \times 64$ volume undergoing spinodal decomposition.  (a) Ground truth.  (b) Our method.  Both (a) and (b) use interlaced acquisition with $K = 16$ and $N_{\theta} = 256$, and a detector size of 363 to ensure full coverage. Top view: volume at the onset of the transformation; bottom view: volume at the end of the transformation.}
\label{fig:3D_recon}
\end{figure}

Figure~\ref{fig:3D_recon} illustrates reconstruction results for a 256$\times$256$\times$64  volume undergoing spinodal decomposition.

\vspace{0.5cm}

\section{Discussion}

From an optimization perspective, our INR-based reconstruction method shares conceptual similarities with classical model-based iterative reconstruction (MBIR) approaches, as it incorporates a data fidelity term derived from the forward model, combined with explicit spatial and temporal regularization. In the case of TIMBIR, for example, the qGGMRF prior acts similarly to a total variation (TV) regularization~\cite{TIMBR}. However, unlike traditional MBIR techniques that operate on discretized image grids, our approach leverages a continuous and implicit neural representation, introducing a learned inductive bias and enabling flexible resolution handling as well as parameter sharing across the temporal dimension.

The model selection strategy used in our experiments, based on tracking the mean residual between primal and dual variables, $\mathbf{x} - \mathbf{q}$ is likely suboptimal. While this criterion promotes internal consistency within the ADMM framework, we observed that it can lead to slight degradations in PSNR and SSIM at later iterations, missing the best model. This suggests that the metrics reported here are conservative and that more principled model selection strategies should be explored in future work.

Finally, although our evaluation is conducted on simulated data, which provides a controlled ground truth for benchmarking, validation on real experimental acquisitions remains an important next step to assess robustness and practical applicability under experimental conditions.

\section{Conclusion}

This work demonstrates the feasibility and effectiveness of leveraging INRs for dynamic XCT reconstruction under interlaced acquisition schemes. The proposed model outperforms traditional baselines across a variety of scenarios, while offering a modular and flexible architecture capable of accommodating detector non-idealities such as ring artifacts. These results highlight the potential of INRs as a compact and versatile framework for tomographic imaging in complex, time-resolved settings, such as in situ monitoring of alloy solidification.

Future work will focus on improving computational efficiency and architectural parallelism to ensure scalability on large datasets, including support for multi-GPU and distributed execution environments.

\section{Acknowledgement}
This work is funded by MIT Startup and the Nuclear Engineering University Program with contract number"DE-NE009491. E.J. acknowledges the support from the John Hardwick Career Development Chair, and M.B. acknowledges support from the MIT-INSTN (CEA) internship program.

\section{Data and code availability}
\label{sec:code}
The code and data used in this study are available at https://github.com/JossouResearchGroupMIT/Interlaced-INR-XCT.

\bibliography{references}

\end{document}